\documentclass{PoS}
\usepackage{array}
\usepackage{graphicx}
\usepackage{wrapfig}
\usepackage{bm}

\def\dt{\delta\tau}		
\def\dH{\delta H}		
\def\DH{\Delta H}		
\def\trjlen{\tau}		
\def\defn{\equiv}
\def\hats{\hat S}		
\def\hatt{\hat T}		
\def\SST{\{S,\{S,T\}\}}
\def\TST{\{T,\{S,T\}\}}
\def\TTTST{\{T,\{T,\{T,\{S,T\}\}\}\}}
\def\STSST{\{\{S,T\},\{S,\{S,T\}\}\}}
\def\TSSST{\{T,\{S,\{S,\{S,T\}\}\}\}}
\def\TTSST{\{T,\{T,\{S,\{S,T\}\}\}\}}
\def\SSSST{\{S,\{S,\{S,\{S,T\}\}\}\}}
\def\STTST{\{\{S,T\},\{T,\{S,T\}\}\}}
\def\O{{\cal O}}		
\def\rational#1#2{{\mathchoice{\textstyle{#1\over#2}}%
  {\scriptstyle{#1\over#2}}{\scriptscriptstyle{#1\over#2}}{#1/#2}}}
\def\half{\rational12}		
\def\dd#1#2{{\mathchoice{d#1\over d#2}%
  {d#1\over d#2}%
  {d#1\!/\!d#2}%
  {d#1\!/\!d#2}}}		
\def\pdd#1#2{{\mathchoice{\partial#1\over\partial#2}%
  {\partial#1\over\partial#2}%
  {\partial#1\!/\!\partial#2}%
  {\partial#1\!/\!\partial#2}}}	
\def\ddq{\pdd{}q}		
\def\ddp{\pdd{}p}		
\newif\ifdragt \dragtfalse
\def\rational#1#2{{\mathchoice{\textstyle{#1\over#2}}%
  {\scriptstyle{#1\over#2}}{\scriptscriptstyle{#1\over#2}}{#1/#2}}}
\def\half{\rational12}		
\def\vec#1{{\bm #1}}		
\def\defn{\equiv}		
\def\M{{\cal M}}		
\def\SU{\mathop{\rm SU}}	
\def\dt{\delta\tau}		
\def\CT#1#2{[#1,#2]}		
\ifdragt
  \def\colonop#1{\mathopen:#1\mathclose:}
  \def\HV#1{\colonop{#1}}	
  \def\PB#1#2{[\colonop{#1,#2}]} 
\else
  \def\HV#1{{\setbox0=\hbox{$#1$}%
     \ifdim\wd0>1em \widehat\vec{{#1}}\else\hat\vec{{#1}}\fi}}
  \def\PB#1#2{\{#1,#2\}}
\fi
\def\sqr#1#2{{\vcenter{\hrule height.#2pt
   \hbox{\vrule width.#2pt height#1pt \kern#1pt
      \vrule width.#2pt}
   \hrule height.#2pt}}}
\def\stpl#1#2{{\vcenter{\hrule height.#2pt
   \hbox{\vrule width.#2pt height#1pt \kern#1pt
      \vrule width.#2pt}}}}

\newlength{\halffigwidth}
\usepackage{subfigure}
\title{Tuning HMC using Poisson brackets}

\ShortTitle{Tuning HMC using Poisson brackets}

\author{M. A. Clark\\
    Center for Computational Sciences, Boston University,\\
    3 Cummington Street, Boston, MA 02215, United States of America\\
    E-mail: \email{mikec@bu.edu}}

\author{
  A. D. Kennedy and P. J. Silva\\
    School of Physics, The University of Edinburgh \& SUPA,\\
    Mayfield Road, Edinburgh, EH9 3JZ, United Kingdom\\
    E-mail: \email{adk@ph.ed.ac.uk, psilva@ph.ed.ac.uk}}

\abstract{We discuss how the integrators used for the Hybrid Monte Carlo (HMC) algorithm not only approximately conserve some Hamiltonian $H$ but exactly conserve a nearby shadow Hamiltonian  \(\tilde H\), and how the difference $\Delta H \equiv \tilde H - H $ may be expressed as an expansion in Poisson brackets. By measuring average values of these Poisson brackets over the equilibrium distribution $\propto e^{-H}$ generated by HMC we can find the optimal integrator parameters from a single simulation. We show that a good way of doing this in practice is to minimize the variance of  $\Delta H$ rather than its magnitude, as has been previously suggested. Some details of how to compute Poisson brackets for gauge and fermion fields, and for nested and force gradient integrators are also presented. }

\FullConference{The XXVI International Symposium on Lattice Field Theory\\
		 July 14-19 2008\\
		 Williamsburg, Virginia, USA}

\begin{document}

\section{Introduction and motivation}

Hybrid Monte Carlo \cite{hmc} is the algorithm of choice to generate dynamical configurations for lattice QCD. This algorithm relies on the introduction of a fictitious momentum for each dynamical degree of freedom, resulting on a Markov chain with a fixed point $\exp(-H(q,p))$ where the Hamiltonian is $H=\frac{1}{2}p^2+S(q)=T(p)+S(q)$; ignoring momenta $p$, we get the desired distribution $exp(-S(q))$.

The HMC Markov chain alternates two Markov steps: \emph{Molecular Dynamics Monte Carlo}, which consists of a reversible volume-preserving approximate Molecular Dynamics trajectory of  $\tau/\delta\tau$ steps followed by a Metropolis accept/reject test with acceptance probability $\min(1,e^{-\delta H})$; and \emph{Momentum refreshment} from a Gaussian heatbath  $P(p) \propto e^{-p^2/2}$.

\subsection{Symplectic integrators}

Symmetric symplectic integrators form a  large class of reversible
and volume-preserving integrators. 
The idea of a \emph{symplectic integrator} is to write the evolution
operator as \(\exp\left(\tau\dd {}t\right) = \exp\left(\tau\left\{ \dd pt \ddp
+ \dd qt\ddq\right\}\right) \defn e^{\tau\hat H}\) where the \emph{ Hamiltonian vector field} \[\hat H = -\pdd Hq\ddp + \pdd Hp\ddq = -S'(q)\ddp + 
T'(p)\ddq \defn \hats + \hatt.\]

We now make use of the Baker--Campbell--Hausdorff (BCH) formula, which tells us
that the product of exponentials in any associative algebra can be written as
\(\ln(e^{A/2}e^Be^{A/2}) - (A + B) = \rational1{24}\left\{\strut [A,[A,B]] -
2[B,[A,B]]\right\} +\cdots\) where all the terms on the right hand side are
constructed out of commutators of \(A\) and \(B\) with known coefficients.  We
find that for an $STS$ integrator with step size \(\dt\) the
evolution operator for a trajectory of length \(\trjlen\) may be written as
\begin{eqnarray*}
  U_{\mbox{\tiny STS}}(\dt)^{\trjlen/\dt} 
  &=& \left(e^{\half\dt\hats} e^{\dt\hatt}
    e^{\half\dt\hats}\right)^{\trjlen/\dt} \\
  &=& \left(\exp\left[(\hatt + \hats)\dt 
    - \rational1{24}\left(\strut[\hats, [\hats,\hatt]] 
    + 2[\hatt,[\hats,\hatt]]\right) \dt^3
    + \O(\dt^5)\right]\right)^{\trjlen/\dt} \\
  &=& \exp\left[\trjlen\left(\hatt + \hats
    - \rational1{24}\left(\strut[\hats, [\hats,\hatt]]
    + 2[\hatt, [\hats,\hatt]]\right) \dt^2
    + \O(\dt^4)\right)\right].
\end{eqnarray*}

\subsection{Shadow Hamiltonians and integrator tuning}

For every symplectic integrator there is a \emph{shadow Hamiltonian} \(\tilde
H\) that is exactly conserved; this may be obtained by replacing the
commutators \([\hats,\hatt]\) in the BCH expansion with the \emph{Poisson
bracket} \(\{S,T\} \defn \displaystyle \pdd Sp\pdd Tq - \pdd Sq\pdd
Tp\)~\cite{inexact:2007}.  For example, the integrator above exactly conserves
the shadow Hamiltonian \(\tilde H_{STS} \defn T + S - \rational1{24}\left(\strut\SST
+ 2\TST \right)\dt^2 + \O(\dt^4)\).  We now make the simple observation that
all symplectic integrators are constructed from the same Poisson brackets  
(which are extensive quantities).
We therefore propose to measure the
average values of the Poisson brackets \(\left\langle\SST\right\rangle\) and
\(\left\langle\TST\right\rangle\) 
over a few equilibrated trajectories at the
parameters of interest and then optimize the integrator (by adjusting the step
sizes, order of the integration scheme, integrator parameters, number of
pseudofermion fields, etc.~\cite{lat2007, Clark:2006fx,forcrand:2006} offline) so as to
minimize the cost.  

As a simple example consider the $STSTS$ integrator \newline
\(U_{\mbox{\tiny{STSTS}}}(\dt)^{\tau/dt} = \left(e^{\alpha\hats\dt}
e^{\half\hatt\dt} e^{(1-2\alpha)\hats\dt} e^{\half\hatt\dt} e^{\alpha
\hats\dt}\right)^{\tau/dt}\) 
whose shadow Hamiltonian is 

\begin{equation}
\tilde H_{STSTS} = H_{STSTS} +
\left(\frac{6\alpha^2 - 6\alpha + 1}{12} \SST + \frac{1 - 6\alpha}{24} \TST
\right)\dt^2 + \O(\dt^4).
\label{shadow:pqpqp}
\end{equation}

 Here we cannot completely eliminate the coefficient of the \(O(\dt^2)\) 
contribution as we only have one free parameter $\alpha$, but we can attempt to minimise the cost by adjusting $\alpha$ given the mean values \(\left\langle\SST\right\rangle\) and \(\left\langle\TST\right\rangle\). Na{\"i}vely we could try to minimize the coefficient of \(\dt^2\) in (\ref{shadow:pqpqp}), but we will see below that this is not the best thing to do.

\subsection{Force gradient integrators \label{fg}}

Let us consider again the $STSTS$ integrator,
where we set \(\alpha=\frac{1}{6}\) so that the \(\TST\) contribution is
eliminated. The remaining leading order Poisson bracket \(\SST\) depends only
on \(q\), which means that we can evaluate the integrator step
\(e^{\widehat{\SST}\dt^3}\) explicitly,

\begin{displaymath}
U_{FG}(\dt)=e^{\frac{\dt}{6}\hat{S}}e^{\frac{\dt}{2}\hat{T}}
e^{\frac{48\dt\hat{S}-\widehat{\SST}\dt^3}{72}}
e^{\frac{\dt}{2}\hat{T}}e^{\frac{\dt}{6}\hat{S}}.
\end{displaymath}

The force for this integrator step involves second derivatives of the action, and
therefore they are called Hessian or force gradient integrators
\cite{chin:2000,omelyan:2002}.
By putting such an integration step into a
multistep integrator we can eliminate all the leading \(\O(\dt^2)\) terms in
\(\DH\), as we can see from the  corresponding shadow Hamiltonian:

\begin{eqnarray*}
\tilde{H}_{FG} = T + S - \rational{\dt^4}{155520} &\Big(& 41\;\SSSST + 36\;\STSST\\
	&&+     72\;\STTST + 84\;\TSSST   \\
	&&+     126\;\TTSST + 54\;\TTTST \Big). \\
\end{eqnarray*}

Note that the coefficients of the leading order correction in the shadow Hamiltonian are approximately two orders of magnitude smaller than the corresponding coefficients in the Campostrini integrator \cite{lat2007, campostrini89a, creutz89a}.

\subsection{Nested integrators}

If it is much cheaper to evaluate the force for one part of the action, such as
the pure gauge part, we can use a nested integrator with a small step
size for the inner cheap part.  One might expect that one could then tune
the outer part without reference to the cheap part, but this is not the
case.

Let the Hamiltonian be \(H = \frac{\pi}{2}^2 + S_1 + S_2\) with \(\|S_2\| \ll
\|S_1\|\) and consider a nested integrator with a composite step of the form
\(U(\dt) = \exp\frac{\hats_2\dt}{2} \left(\exp\frac{\hats_1\dt}{2m}
\exp\frac{\hatt\dt}{m} \exp\frac{\hats_1\dt}{2m} \right)^m \exp\frac{\hats_2
\dt}{2}\).  For the inner integrator the BCH formula tell us that \(\left(\exp
\frac{\hats_1 \dt}{2m} \exp\frac{\hatt\dt}{m} \exp\frac{\hats_1\dt}{2m}
\right)^m\) may be written as \[\exp\left((\hats_1+\hatt)\dt + \Big(\alpha
\CT{\hats_1}{\CT{\hats_1}{\hatt}} + \beta \CT{\hatt}{\CT{\hats_1}{\hatt}}
\Big) \frac{\dt^3}{m^2} + \O(\dt^5)\right)\] with \(\alpha=-\frac1{24}\) and
\(\beta=\frac1{12}\).  Applying the BCH formula again leads to
the shadow Hamiltonian
\begin{eqnarray*}
\tilde H = H &+& \biggl(\alpha
    \PB{\hats_2}{\PB{\hats_2}{\hatt}} + \beta \PB{\hats_1}{\PB{\hats_2}
    {\hatt}} + \beta \PB{\hatt}{\PB{\hats_2}{\hatt}} \\
  && \qquad + \frac1{m^2}\Bigl( \alpha
    \PB{\hats_1}{\PB{\hats_1}{\hatt}} + \beta \PB{\hatt}{\PB{\hats_1} {\hatt}} 
    \Bigr)\biggr) \dt^2 + \O(\dt^4).
\end{eqnarray*}
Observe that the Poisson bracket \(\PB{\hats_1}{\PB{\hats_2}{\hatt}}\) depends
on the cheap action \(S_1\) but is not supressed by any inverse power of $m$;
it is therefore still necessary to measure this quantity in order to optimize the
integrator. 

\section{Computing Poisson brackets}

\subsection{Gauge fields}

We must construct the Poisson brackets for gauge fields, where the field
variables are constrained to live on a group manifold. To do this we need to
use some differential geometry \cite{lat2007}.
In order to construct a Hamiltonian system on such manifold we need not only a
Hamiltonian function but also a fundamental closed 2-form \(\omega\).  On a Lie
group manifold this is most easily found using the globally defined
\emph{Maurer--Cartan} forms \(\theta^i\) that are dual to the generators and
satisfy the relation \(d\theta^i = -\half c^i_{jk} \theta^j\wedge \theta^k\),
where \(c^i_{jk}\) are the structure constants of the group.  We choose to
define \(\omega \defn -d\sum_i \theta^ip^i = \sum_i(\theta^i\wedge dp^i -
p^id\theta^i) = \sum_i (\theta^i \wedge dp^i + \half p^i c^i_{jk} \theta^j
\wedge\theta^k)\): using this fundamental 2-form we can define a Hamiltonian
vector field \(\hat A\) corresponding to any 0-form \(A\) through the relation
 $ dA ( \vec{x}) = \omega(\hat A,\vec x)$ for all vector fields $\vec{x}$.

For a Hamiltonian of the form \(H=S+T\) we find that the leading Poisson
brackets that appear in the shadow Hamiltonian for a symmetric symplectic
integrator are \(\SST = e_i(S)e_i(S)\) and \(\TST = -p^ip^je_ie_j(S)\) where
the \(p^i\) are the momentum coordinates and the \(e_i\) are linear
differential operators satisfying \(e_i(U) = T_iU\) for gauge fields
\(U\in\SU(n)\) with generators~\(T_i\).

\subsection{Fermions}

Consider the Wilson pseudofermionic action \(S=\phi^\dagger \M^{-1} \phi\), and recall that the \(e_i\) are linear differential operators,
thus \(e_i (S) = -\phi^\dagger \M^{-1} e_i (\M) \penalty-1000 \M^{-1}\phi\),
and 

\begin{displaymath}
p^ip^je_ie_j (S) = p^ip^j \phi^\dagger \M^{-1} \left[ 2\,e_i (\M) \M^{-1}
e_j (\M) - e_ie_j (\M) \right ] \M^{-1} \phi.
\end{displaymath}

  \(e_i(\M)\) is straightforward
to evaluate given the linearity of the Wilson--Dirac operator in the gauge
field: we just use Leibniz rule and then replace the gauge field \(U\)
by ~\(T_i U\).

\section{Results}

\subsection{Shadow Hamiltonian and Poisson brackets}

\begin{figure}[!t]
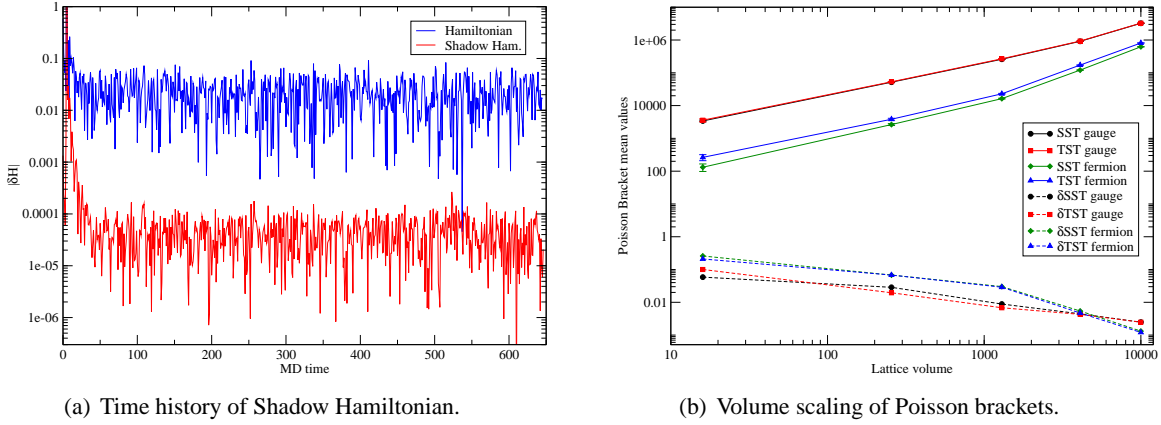

  \subfigure[Time history of Shadow Hamiltonian.]{
  \begin{minipage}[b]{0.45\textwidth}
    \centering
    \includegraphics[origin=c, angle=0, width=\halffigwidth]{deltah.eps}
  \end{minipage} } \hfill
  \subfigure[Volume scaling of Poisson brackets.  ]{
  \begin{minipage}[b]{0.45\textwidth}
    \centering
    \includegraphics[origin=c, angle=0,width=\halffigwidth]{vol_scal_delta.eps}
  \end{minipage} }
\caption{Shadow Hamiltonian and Poisson brackets.}
\label{shpb}
\end{figure}

The blue curve in the first plot of figure \ref{shpb} shows how \(\log_{10}|\dH|\equiv \log_{10}|H_f-H_i|\)
behaves as a function of MD time, compared with the red curve \(\log_{10}
|\delta\tilde H|\) for the shadow Hamiltonian up to leading non-trivial order
in \(\dt\).  The simulation uses the $STSTS$ integrator with  $\alpha=0.24$ and  $\delta \tau=0.1$ for Wilson gauge and fermion actions. This demonstrates that 
the shadow Hamiltonian is indeed conserved.

The second graph on figure \ref{shpb} shows how several different 
Poisson brackets and their fluctuations depend on the lattice size.  
As expected the Poisson brackets are more-or-less extensive (they grow 
as \(L^4\)); the statistical fluctutations in the Poisson brackets 
are also shown, and they fall as \(L^{-2}\) relative to the  mean 
values as expected.

\subsection{How to tune an integrator?}

\begin{wrapfigure}{r}{0.45\textwidth}
\vspace*{-0.8cm}
   \centering
   \includegraphics[width=0.4\textwidth]{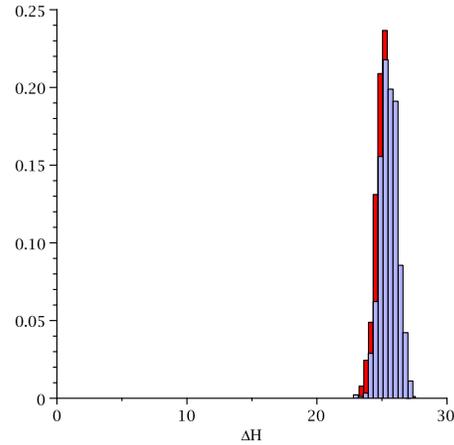}
   \caption{\label{hist1m} Histogram of $\Delta H$ at the start (blue) and end (red) of the trajectories. }
\vspace*{-0.5cm}
\end{wrapfigure}

We are concerned in minimizing the cost of HMC; in our case, this corresponds 
to maximizing the step size $\dt$ while maintaining a reasonable acceptance 
rate. The first step to this goal is to find the integrator parameters 
that maximize the acceptance rate for a given value of $\dt$. Here we are 
going to discuss results for the $STSTS$ integrator described above, trying 
to find the optimal value for $\alpha$.

Omelyan \textit{et al.} \cite{omelyan:2002} proposed that one should 
minimize $\langle\Delta H^2\rangle\equiv\big\langle(\tilde{H}-H)^2\big\rangle$,
 as this makes $\tilde{H}$ as close to $H$ as possible. However, the amount
 by which  $\Delta H$ varies over the equilibrium distribution 
$\propto e^{-H}$ turns out to be considerably smaller than the values 
of $\Delta H$ itself. Therefore, it seems more reasonable to minimize 
$\mbox{Var}(\Delta H)$, the variance of $\Delta H$ over this equilibrium distribution.

Indeed, figure \ref{hist1m} verifies that $\langle \Delta H \rangle \gg \sqrt{\mbox{Var}(\Delta H)}$. If we assume that $H_f$ and $H_i$ are selected independently from their equilibrium distributions, which is a goal of HMC,
 $\langle \Delta H \rangle \gg \langle \delta H \rangle$ as figure \ref{hist1m} also verifies. We can also conclude that the initial and final distributions seem to be equivalent --- of course, $H_f$ is not distributed according to the equilibrium distribution as $H_i$ is, but its distribution does not differ significantly.

\begin{figure}[t]
\vspace*{-0.3cm}
  \subfigure[$STSTS$ integrator.]{
  \begin{minipage}[b]{0.45\textwidth}
    \centering
    \includegraphics[origin=c, angle=0, width=0.85\halffigwidth]{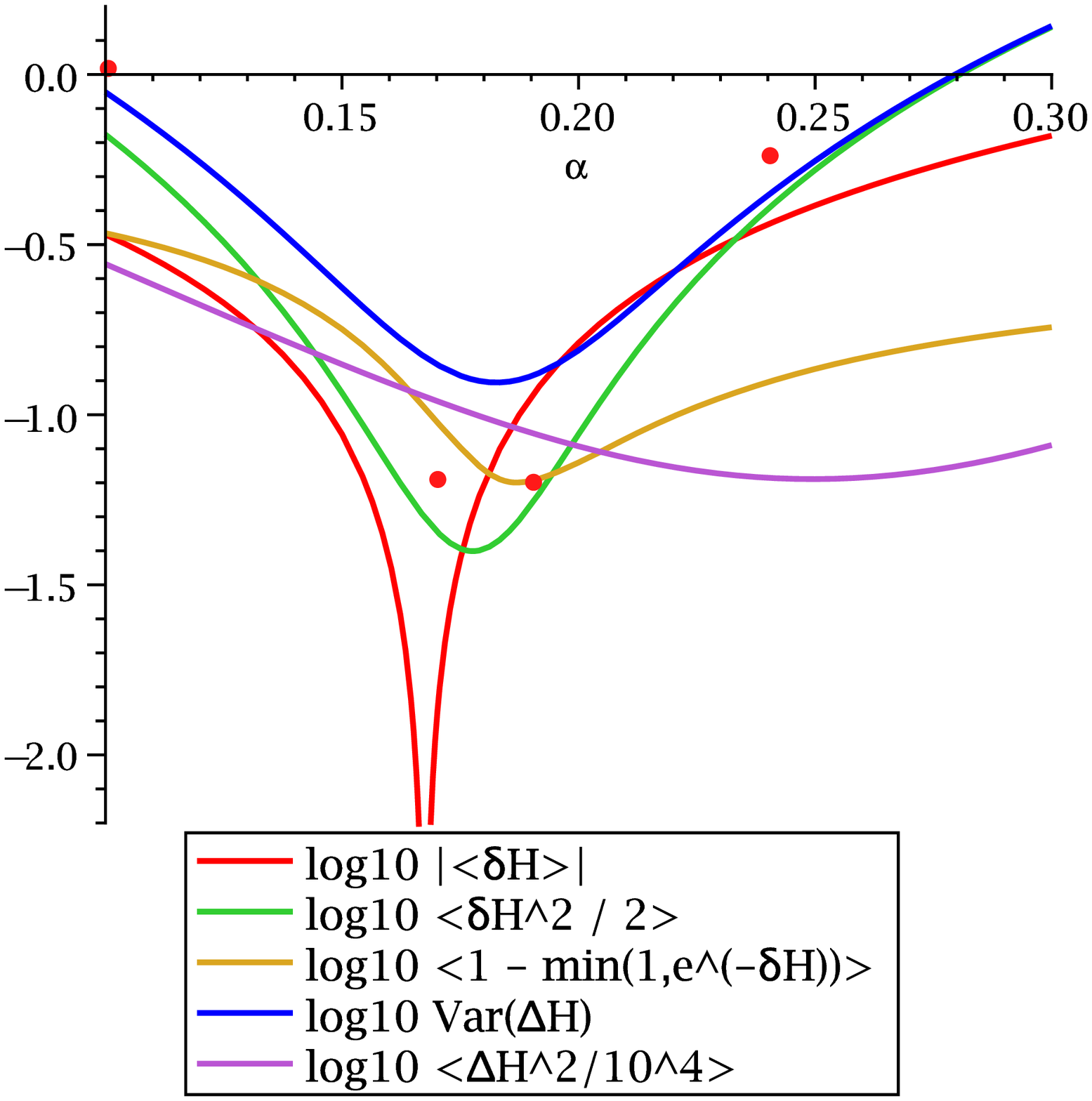}
  \end{minipage} \label{tuning:1}} \hfill
  \subfigure[Two level $STSTS$. ]{
  \begin{minipage}[b]{0.45\textwidth}
    \centering
    \includegraphics[origin=c, angle=0,width=0.85\halffigwidth]{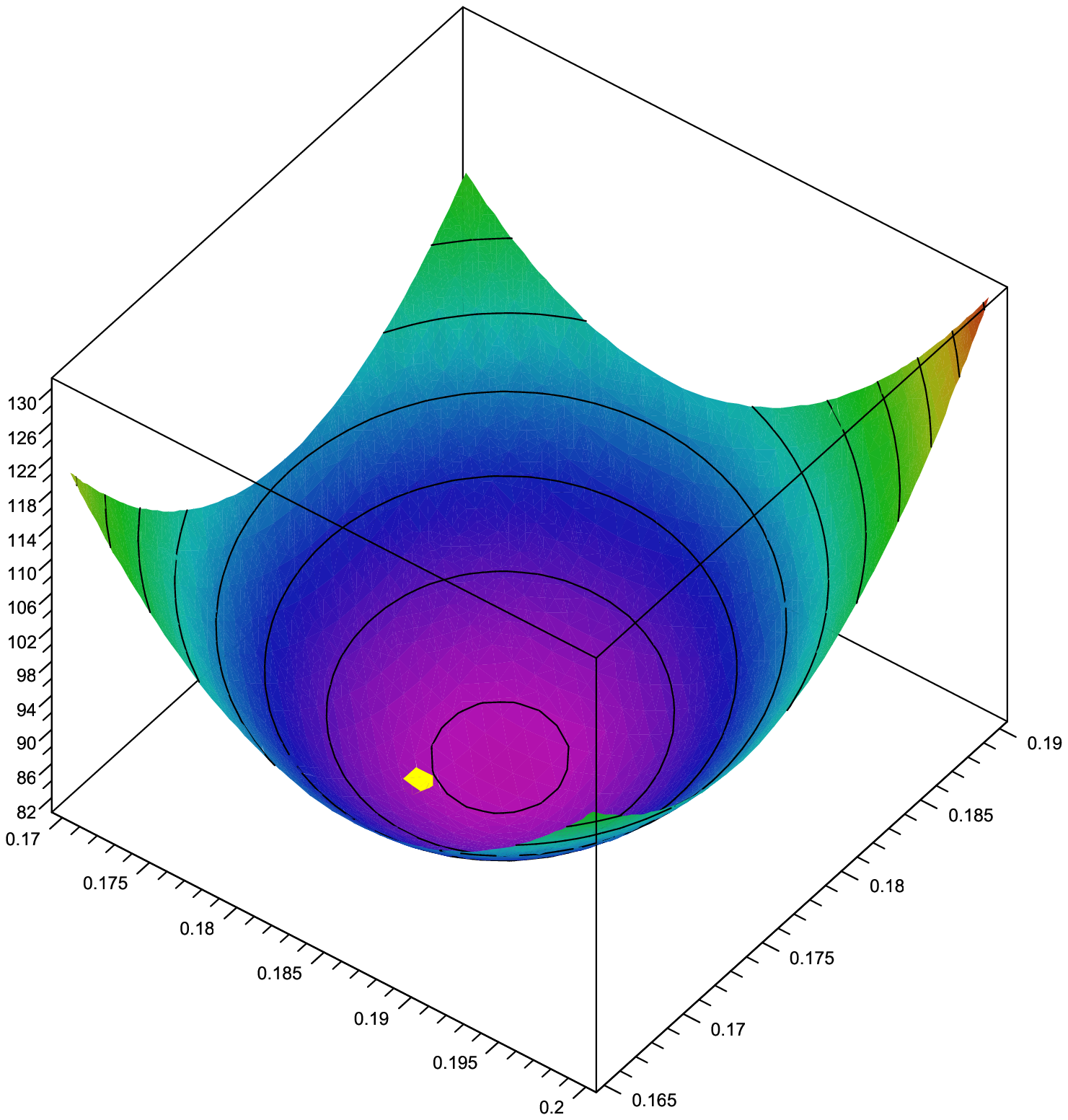}
  \end{minipage} \label{tuning:2}}
\caption{\label{tuning} Tuning plots.}
\end{figure}

\begin{wraptable}{l}{0.45\textwidth}
\vspace*{-0.3cm}
\begin{center}
\begin{tabular}{l@{\hspace*{0.3cm}}l}
\hline
Function   & $\alpha_{min}$ \\
\hline
$|\langle \delta H \rangle |$ & 0.16749  \\
$\langle \delta H^2/2 \rangle$  & 0.17765  \\
$ \mbox{Var}(\Delta H)$         & 0.18260  \\
$\langle 1 - \mbox{min}(1,e^{-\delta H}) \rangle $ & 0.18664  \\
\hline
$\langle \Delta H^2 \rangle $  & 0.24952\\
\hline
\end{tabular}
\caption{\label{tuntable} Optimal values for $\alpha$.}
\end{center}
\vspace*{-0.3cm}
\end{wraptable}

In figure \ref{tuning:1}, we see plots of several quantities, besides 
$\mbox{Var}(\Delta H)$, we could minimize to optimize the integrator. 
The curves were computed using the Poisson brackets computed at 
$\alpha=0.24$, whereas the red points are measurements of 
$\langle \delta H^2/2 \rangle$  at different $\alpha$ values. The good 
agreement between the measured and predicted location of the minimum 
gives us confidence that we can find the correct behaviour of the quantities 
of interest by measuring the Poisson brackets at a single value of the 
integrator parameters. 

In table \ref{tuntable} we can see the optimal $\alpha$ values for the 
quantities considered. We see that the minima for 
$|\langle \delta H \rangle |$, $\langle \delta H^2/2 \rangle$ 
and $\langle 1 - \mbox{min}(1,e^{-\delta H}) \rangle $ 
are close to the minimum of $ \mbox{Var}(\Delta H)$.

Figure \ref{tuning:2} shows similar results for tuning the parameters for a
dynamical fermion computation on a \(8^4\) lattice with a Wilson gauge action
with \(\beta=5.6\) and Wilson fermions with \(\kappa=0.1575\). Here we minimize  \(\langle \dH^2\rangle\).  We used a two
level $STSTS$ integrator with two gauge steps per fermion step, and a trajectory
length of one.  The yellow point shows values of the \(\alpha\) parameters at
which the Poisson brackets were measured.

\subsection{Force gradient integrators}

\begin{figure}[t]
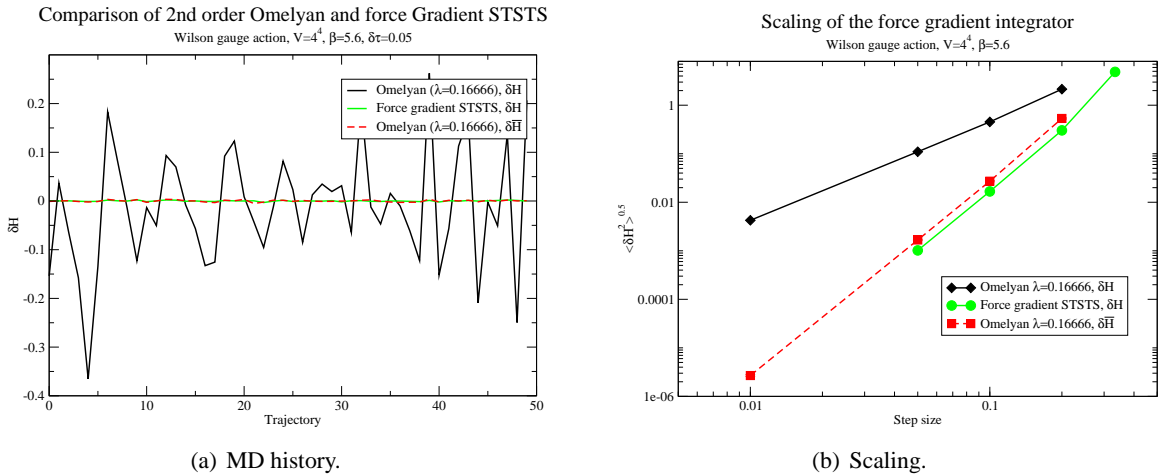


  \subfigure[MD history.]{
  \begin{minipage}[b]{0.45\textwidth}
    \centering
    \includegraphics[origin=c, angle=0, width=\halffigwidth]{dh.eps}
  \end{minipage} \label{fgfig:h} } \hfill
  \subfigure[Scaling. ]{
  \begin{minipage}[b]{0.45\textwidth}
    \centering
    \includegraphics[origin=c, angle=0,width=\halffigwidth]{scale.eps}
  \end{minipage}\label{fgfig:s} }
\caption{\label{fgfig} Results for the $STSTS$ force gradient integrator 
(green data), compared with data for second order Omelyan integrator (black 
data). Note also the data for the shadow Hamiltonian of Omelyan integrator 
(red data).}
\end{figure}

In this subsection, we show results for the force gradient integrator 
defined in section \ref{fg}, obtained with the Wilson gauge action at 
$\beta=5.6$ on a $4^4$ volume, comparing with a second order Omelyan 
integrator (figure \ref{fgfig}). Note that the scaling for the force 
gradient integrator (black data in figure \ref{fgfig:s}) is much better 
than for the Omelyan integrator (green data in figure \ref{fgfig:s} ).

\section{Conclusions}

We have shown that a good strategy to optimize HMC integrators is to minimize 
the variance of $\Delta H$ over the equilibrium distribution $e^{-H}$, rather 
than minimizing $|\Delta H|$ itself, as was previously proposed. 
We have outlined how the Poisson brackets required to compute $\Delta H$ may 
be evaluated for gauge theories and systems with dynamical fermions. We have 
also carried out initial investigations with nested integrators and force 
gradient integrators. 
We hope to present more details of our results, and data for more realistic 
computations soon.

\acknowledgments

This work was supported in part by  PPARC/STFC grant
PP/D000238/1 and NSF grant PHY-0427646.  
P. J. Silva acknowledges support from FCT via grant SFRH/BPD/40998/2007.

\end{document}